\documentclass[pra,preprint]{revtex4}
\usepackage{epsfig}
\usepackage{graphicx}
\usepackage{epstopdf}
\newcommand{\eq}[1]{Eq.(\ref{#1})}
\newcommand{\be}[1]{\begin{equation}\label{#1}}
\newcommand{\ee}{\end{equation}}
\newcommand{\bea}{\begin{eqnarray}}
\newcommand{\eea}{\end{eqnarray}}

\begin{document}
\title{Long-time and unitary properties of semiclassical initial value 
representations}
\author{C.~Harabati$^{1}$, J.~M. Rost$^{1}$, and F.~Grossmann$^{2}$}
\affiliation{$^{1}$Max-Planck-Institute for the Physics of Complex Systems, 
N\"{o}thnitzer Strasse 38, D-01187 Dresden, Germany\\
$^2$Institute for Theoretical Physics, Dresden University of Technology,
D-01062 Dresden, Germany}
\date{\today}

\begin{abstract}
We numerically compare the semiclassical
``frozen Gaussian'' Herman-Kluk propagator [Chem. Phys. {\bf 91}, 27 (1984)] 
and the ``thawed Gaussian'' propagator put forward recently by 
Baranger et al.\ [J. Phys. A  {\bf 34}, 7227 (2001)] by studying 
the quantum dynamics in some nonlinear one-dimensional potentials. 
The reasons for the lack of long time accuracy and norm conservation in 
the latter method are uncovered. We amend the thawed Gaussian
propagator with a global harmonic approximation for the stability of
the trajectories and demonstrate that this revised propagator is a
true alternative to the Herman-Kluk propagator with similar accuracy.
\end{abstract}
\maketitle
\newpage
\section{Introduction}

The inclusion of quantum effects into classical dynamical calculations
is of fundamental as well as of practical interest. It may, e.\ g., help
in understanding the nature of interference related effects and, 
for systems with many degrees of freedom (DOF), 
it may make quantum calculations possible in the first place.
The most fundamental approach to reach this goal is based on the 
application of the semiclassical approximation to the quantum mechanical 
path integral expression for the time-dependent propagator \cite{FH65}.  
A major drawback of the standard time-dependent semiclassical 
Van Vleck-Gutzwiller formula \cite{VV28,Gu67} is 
the so-called root search problem. One has to find solutions to 
classical boundary value problems, i.\ e.\ classical trajectories connecting
a given initial with a given final position in a specified time. 
This may become numerically cumbersome for longer times, especially if many 
DOF are involved in the classical dynamics, as is usually
the case.

Initial value representations (IVRs) of the semiclassical propagator have 
therefore been the focus of interest for quite some time
\cite{Miller,He81,HK84,He91,Kay1,Kay2,Gross99}. In the IVRs,
root searches are avoided by transforming the semiclassical propagator 
into an integral expression leading to the emergence
of classical initial value problems instead of boundary ones.
Specifically the Herman-Kluk (HK) IVR expression based on
fixed width, i.\ e.\ ``frozen'' Gaussians has been quite successfully 
applied since its first derivation \cite{HK84}. Recently, an alternative 
formulation, based on ``thawed'' Gaussians, has been 
brought back into the limelight in Refs.\ \cite{Bara1} and \cite{Bara2}.
The single trajectory version, for a propagated Gaussian in the
formulation of Ref.\ \cite{Bara1} is equivalent to Heller's 
Gaussian wave packet dynamics \cite{He75}, evolving a Gaussian 
wave packet by using only
its  center trajectory and allowing the width of the Gaussian 
to ``breathe''. The full integral expression based on
thawed Gaussians (i.\ e.\ the multitrajectory version, see also Ref. 
\cite{Bara2}) has already been discussed by Kay \cite{Kay1} 
and has been labelled thawed Gaussian approximation (TGA). 
Rather surprisingly, Kay found first numerical evidence
that the TGA is numerically far inferior as compared to the HK 
frozen Gaussian Approximation (HK FGA). As the main result of this article,
we propose a ``breathing'',  individually determined  from the global
properties of the classical dynamics on  the energy shell which is defined  
for each trajectory by its initial point in phase space, in order to
adjust the shortcomings of TGA.

In the following, we will explicitly examine several examples  
to shed some light on the reason for the
failure  of the original TGA. Simultaneously, it will become clearer why  FGA 
is numerically superior. To set the stage, 
in Sec. II, we briefly review Kay's results \cite{Kay} of a general
semiclassical IVR expression based on coherent states. In Sec. III,
the performance of the two special cases, HK FGA and TGA, is contrasted
for two different anharmonic test potentials. In Sec. IV, the reason
for the TGA failure is 
underlined by a simple improvement before we introduce the global
harmonic approximation
which improves TGA to the level of FGA. Conclusions and an outlook
are given in Sec. V.

\section{Semiclassical Initial value representations}

A general expression for the quantum propagator, subsuming many of the 
previously known semiclassical initial value representations, has been derived
via the principle of stationary phase equivalence by Kay \cite{Kay1}.
He used the fact that the Ansatz as an integral over phase space
has to lead to the Van Vleck-Gutzwiller form, if the integration is
performed in the stationary phase approximation.

Several early semiclassical approaches, e.\ g.\ the ones of Heller \cite{He75}
and Herman and Kluk \cite{HK84} are based on Gaussian wave packets of the form 
\be{coherent-state}
\langle x | g_\gamma(p,q)\rangle=
\left(\frac{{\rm Re}\gamma}{\pi}\right)^{1/4}
\exp\left\{-\frac{\gamma}{2}
(x-q)^2 + \frac{i}{\hbar}p(x-q)\right\}
\end{equation}
in coordinate representation.
In his seminal work \cite{Kay1}, Kay has used these states with arbitrarily 
time-dependent width parameters $\gamma(t)$ in order to construct a general
semiclassical integral expression for the propagator.
For our purposes, we need only the special version without a cross term
between initial and final Gaussian, given by
\be{eq:Kay}
K(x,t;x',0)=\int \frac{dp_i dq_i}{2\pi\hbar}
\langle x | g_{\gamma_1}(p_t,q_t)\rangle R(p_i,q_i,t)\exp\{iS(p_i,q_i,t)/\hbar\}
\langle g_{\gamma_2}(p_i,q_i)|x'\rangle,
\ee
where the pre-exponential factor, 
\be{eq:pre}
R(p_i,q_i,t)=\left[\frac{1}{2\sqrt{{\rm Re}\gamma_1{\rm Re}\gamma_2}}
\left(\gamma_1\frac{\partial q_t}{\partial q_i}
+\gamma_2\frac{\partial p_t}{\partial p_i}
-i\hbar \gamma_1\gamma_2\frac{\partial q_t}{\partial p_i}
-\frac{1}{i\hbar}\frac{\partial p_t}{\partial q_i}
\right)\right]^{1/2}
\ee
is determined by the stability matrix  (or monodromy matrix) elements defined in the matrix
equation,
\be{stability-matrix}
\left( \begin{array}{c}
\delta q_t \\
\delta p_t
\end{array} \right)
=\left(\begin{array}{cc}
\partial q_t/\partial q_i & \partial q_t/\partial p_i \\
\partial p_t/\partial q_i & \partial p_t/\partial p_i 
\end{array} \right)
\left( \begin{array}{c}
\delta q_i \\
\delta p_i
\end{array} \right)
\end{equation}
which connect the initial with the final deviations of trajectories 
$(p_t,q_t)$ in phase space, while the phase is determined by the classical 
action
$
S(p_i,q_i,t)=\int_0^t L dt'
$
being the time integral over the Lagrange function.
 
From the general expression, Eq.\ (\ref{eq:Kay}), the HK approximation 
is recovered 
by setting $\gamma_1=\gamma_2=\gamma$ with a real positive
constant $\gamma$. The explicit form of the  HK FGA prefactor is then 
\be{eq:pre1}
R^{\rm HK}(p_i,q_i,t)=
\sqrt{\frac{1}{{2}}\left(\frac{\partial q_t}{\partial q_i}+
\frac{\partial p_t}{\partial p_i}-i\hbar \gamma\frac{\partial q_t}
{\partial p_i}-\frac{1}{i\hbar \gamma}\frac{\partial p_t}{\partial q_i}
\right)},
\end{equation}
while by setting 
\be{eq:gamma1}
\gamma_1=-\frac{i}{\hbar}
\frac{\partial p_t/\partial q_i+i\hbar\gamma\partial p_t/\partial p_i}
{\partial q_t/\partial q_i+i\hbar\gamma\partial q_t/\partial p_i}
\ee 
again with real positive $\gamma$ 
\cite{remark2} and $\gamma_2=\gamma$, we recover the TGA. 
This semiclassical IVR has recently also been discussed by Baranger 
et al.\ \cite{Bara1,Bara2}. The prefactor turns out to be 
\be{eq:pre2}
R^{\rm TGA}(p_i,q_i,t)=\left(\frac{\gamma}{\rm{Re}\gamma_1}\right)^{1/4}
\left(\frac{\partial q_t}{\partial q_i}
+i\hbar\gamma\frac{\partial q_t}{\partial p_i}\right)^{-1/2}.
\end{equation}

In the following, we will numerically investigate the TGA expression
and compare the results to FGA and full quantum results.
Later on, the general expression (\ref{eq:Kay}) will then also
allow us to search for improvements of the TGA propagator.


\section{Long time accuracy of HK FGA versus TGA} 

For reasons of simplicity, the wave function to be propagated
is chosen to be a Gaussian $\Psi(x,0)=\langle x|g_\gamma(p_0,q_0)\rangle$ 
localized around $(p_0,q_0)$
in phase space and having the same width parameter as the initial Gaussian 
in the integral expression of the propagator in Eq.\ (\ref{eq:Kay})
\cite{remark3}. 
In the following we will compare 
the overlap between the time propagated and the unpropagated wavepacket, 
the so-called autocorrelation function
\be{aut}
{\rm c}(t)=\int_{-\infty}^{\infty} dx \, \Psi^*(x,0)\Psi(x,t),
\end{equation}
for the different semiclassical approaches.

Inserting the propagator from Eq.\ (\ref{eq:Kay}), 
we obtain the semiclassical autocorrelation function 
{\setlength\arraycolsep{2pt}
\begin{eqnarray}
\label{eq:ac}
{\rm c}(t) &=& \int \frac{dq_i\, dp_i}{2\pi\hbar}
\langle g_\gamma(p_0,q_0)|g_{\gamma_1}(p_t,q_t)\rangle 
R(p_i,q_i,t)
\nonumber\\
& &
\exp\left\{iS(p_i,q_i,t)/\hbar\right\}
\langle g_{\gamma_2}(p_i,q_i)|g_\gamma(p_0,q_0)\rangle.
\end{eqnarray}
The overlap integral between the Gaussians can be performed
analytically, whereas the phase space integration in Eq. (\ref{eq:ac}) 
is done numerically by the Monte Carlo integration 
procedure described in Ref. \cite{KHD86}. 
To test the numerical performance of the different methods we chose
two different nonlinear potentials which we now introduce.

First, we employ the Hamiltonian taken by Baranger et al.\ \cite{Bara1} 
to test their integral kernel, 
\be{eq:Bar}
H_{B}=\frac{p^2}{2\mu}+2V_0 e^{-\alpha A}\cosh(\alpha x).
\end{equation}
Throughout the rest of the article, we use only this classical version of the 
Hamiltonian (without Gaussian smoothing)
with (scaled) units, particularly scaled time $t_{\rm B}$, indirectly
defined by the parameters taken from Ref. \cite{Bara1} as $\mu=1$, $V_0=1$, $A=5$,
$\alpha=1$, and a Planck constant of $\hbar=0.05$. The initial 
conditions of the wave packet are $q_0=0$ and $p_0=1$ and  $\gamma=100/9$.
``Smoothing'' of the Hamiltonian by using its coherent state matrix element
leads only to minor changes (not shown) in the results to be presented below.

As the second test bed, we choose the well known Morse
potential with the Hamiltonian 
\be{eq:Morse}
H_{M}=\frac{p^2}{2\mu}+V_0(1-e^{-\lambda x})^2
\end{equation}
which is  a reasonable
model for a chemical bond. The parameters  
taken from Ref. \cite{He91} as $\lambda=0.08$, $V_0=30$, which 
together with $\mu=1$ define a scaled time $t_{\rm M}$.
As initial conditions for the wavepacket we define
$q_0=p_0=0$ and  $\gamma=12$.

In  Fig.\ \ref{fig:ac+norm}, we compare different semiclassical IVR results
(i.\ e.\, using the full integral expression) to 
the quantum result, calculated by the split operator method, for the
two model potentials. For the Baranger potential, we have calculated
the autocorrelation function 
up to scaled times of 110, as displayed in  Fig.\ \ref{fig:ac+norm}(a).   
In accord with the decay of the norm in Fig.\ \ref{fig:ac+norm}(c),
also the autocorrelation
function shows a strong decay, not present in the full quantum curve.
We state explicitely that the decay of the norm for TGA and also 
the discrepancy of the norm
from unity in the HK FGA case are not in contradiction to the fact
that both methods are unitary in the (analytical) stationary phase sense
(see e.\ g.,\ Ref. \cite{He86} for the HK FGA case).

Nevertheless, for the Baranger potential, the quality of the 
numerical TGA results is
disappointing. For the very short times (up to 10 in scaled units)
investigated in Ref. \cite{Bara1}, even the one trajectory (kernel) version
of TGA, i.\ e.,\ Heller's Gaussian wave packet dynamics, does better
than the full integral version (not shown). 
This method is, however, not suited for longer
times due to the nonlinearity of the potential. 
In contrast, as can also be observed in  Figs.\ \ref{fig:ac+norm}(a)
and \ref{fig:ac+norm} (c), 
the HK FGA approach gives a reasonable agreement with the full
quantum solution.

The same kind of discrepancies of TGA from the quantum split operator 
calculation 
as for the Baranger potential can be observed in the case of the 
Morse potential. The corresponding results are given in 
Figs. \ref{fig:ac+norm} (b) and \ref{fig:ac+norm} (d). 
As has been noted already in \cite{Gross99}, for the Morse
potential, the HK FGA approximation  represents the fine details of
the full quantum result very well. 
The multitrajectory TGA expression, however, already fails again at 
intermediate times.

\section{Improvements of the TGA}

With the comparison of the different results at hand, we have the means to 
find the reason for the poor performance of the TGA expression.

\subsection{Generalized TGA}

The reason for the disappointing TGA results becomes clear by taking a
closer look at the time-dependent width parameter $\gamma_1$. To this
end, we concentrate on the Hamiltonian of \eq{eq:Bar} with the 
Baranger potential.
As can be seen in the lower panel of Fig.\ \ref{fig:gamma_1},
the real part of the complex width parameter is almost always
close to zero for longer times. This leads to a strongly
oscillating exponent in the TGA integral expression. Although the
Monte Carlo results which we present are converged, the strong
oscillation of the integrand leads to the rapid deterioration of the results.
The reason for the satisfactory behavior of the HK FGA in the
cases we investigated is obviously
the fixed real width parameter which by definition does not decay.

This observation opens up a possibility to improve on the standard TGA
expression in a simple way. 
If one could render the width parameter's ``destructive influence''
on the results less dramatic, this would certainly ameliorate the results.
An easy way to achieve this goal is by taking (higher-order) roots
of the width parameter $\gamma_1$. This is done in the upper two
panels of Fig.\ \ref{fig:gamma_1}, where in panel (b)-the square root
-and in panel (a) the fourth root of the width parameter in panel (c)
is taken. Clearly, the decay of the real part is less dramatic.
The effect that this change has on the autocorrelation function
can now be checked by inserting the new width parameter into the
general integral expression of Eq.\ (\ref{eq:Kay}) and calculating
the resulting prefactor. The emerging propagator is then still
semiclassical in the sense of stationary phase equivalence. The numerical
results are shown in the upper panels of Fig. \ref{fig:bara}.
Already in the case of the fourth root (see panel Fig. \ref{fig:gamma_1} (a)), 
the generalized TGA result is almost as good as the HK FGA.
By making the final Gaussians narrower in width, we have reached
an intermediate goal. TGA can now better capture the real quantum dynamics.

There is, however, one further aspect to this improvement.
The issue whether the generalized TGA can do as well as HK FGA
with fewer trajectories. To study this, we have investigated the convergence
of the Monte Carlo integration for FGA, the two different generalized TGAs,
and the standard TGA.
For the Baranger potential example, all methods require on the order
of several times $10^3$ trajectories to achieve convergence.
The HK FGA needs a slightly higher number of trajectories than standard TGA.
Taking roots of higher and higher order makes the width almost constant 
and, therefore, the TGA propagator approaches the FGA. This also
leads to more trajectories which are required for convergence.

Besides taking roots of the width parameter, there is yet another 
way to improve the TGA, making it a true alternative to the FGA.
This approach will be studied in the remainder of this article.

\subsection{Global harmonic approximation}

A starting point for an improvement of the TGA,
 different in spirit from the one presented above,
is based on the fact that all {\it time-dependent} 
semiclassical approaches become exact for the case of the harmonic
oscillator. The TGA is especially favorable for the case of a harmonic
potential, whose frequency $\omega$ does not coincide with the width
parameter of the Gaussians used in the integral expression for the
propagator. Already the single trajectory Gaussian wave packet dynamics
gives the exact analytical solution (the so-called squeezed state)
which breathes (for $\omega\neq \hbar\gamma/\mu$)
according to the time-dependent width parameter
\be{eq:ho}
\gamma_1=-\frac{i}{\hbar}
\frac{-\mu\omega\sin(\omega t)+i\hbar\gamma\cos(\omega t)}
{\cos(\omega t)+i\hbar\gamma/(\mu\omega)\sin(\omega t)}.
\ee
The individual Gaussians of TGA
breathe in the same way as the solution does, allowing the convergence 
of the numerical results to be
achieved with fewer trajectories than in the HK FGA case, also explaining
 the remarks at the end of the previous subsection. Is this observation
of any use for realistic non-harmonic potentials?

The answer is yes and we want to demonstrate this for the case of the
Morse oscillator with the parameters given in the previous section. 
For the bound motion in any potential we can define an action variable 
\be{eq:action}
I=\frac{1}{2\pi}\int_C p dq=\frac{1}{2\pi}\int_C \sqrt{2\mu(E-V)}dq
\ee
by integrating over a closed contour $C$ in phase space.
The Hamiltonian does not depend on the corresponding angle variable
and can be expressed as a function of $I$.
Its derivative, with respect to the action variable, then defines a frequency
\be{eq:freq}
\omega=\frac{\partial H}{\partial I}
\ee
which in anharmonic potentials depends on energy and, in the simple one 
DOF case, can be determined analytically.
The idea of the improvement of TGA is now to use the oscillation
frequency given above as the one of
a ``fictitious'' harmonic oscillator and let every Gaussian breathe
with the corresponding $\gamma_1$. While all stability matrix
elements for the calculation of the semiclassical propagator
are taken to be the harmonic ones, the trajectories themselves
are propagated using the full nonlinear force field. 
For every trajectory, the approximation is a {\it global} harmonic 
approximation, with respect to the stability information. 
Note that the present approach  is different from the
the so-called nonlinear Gaussian wave packet 
dynamics, introduced by Tomsovic and Heller
\cite{TH93} which uses a {\it local} expansion of the potential up
to second order (as Heller's original Gaussian wave packet dynamics) 
and is not an  IVR formulation but a root search method in spirit. 

We are now using the global harmonic approximation for the width 
parameter and also
the prefactor in Eq.\ (\ref{eq:ac}) to calculate the autocorrelation 
function in the Morse oscillator case. The results are 
shown in Fig.\ \ref{fig:morse}. A dramatic improvement toward the HK 
FGA result can be observed. The reason is that now 
$\gamma_1$ behaves as in the harmonic case, i.\ e.\
it does not decay to zero for later times,
thus leading to reasonable results.
We found that the same kind of improvement toward
HK is also present for wave packets
with different initial conditions (not shown).

\section{Conclusions}
Based on numerical studies for two different anharmonic potentials
(Baranger and Morse),
we have identified the decay of the width parameters of the Gaussians in time
as the reason for the failure of the TGA to provide a good semiclassical 
propagator. 
Following this notion, we have introduced the global harmonic approximation 
which mimics
for each trajectory its stability properties in time according to those of a 
harmonic oscillator.
Its frequency is defined through an action-angle analysis of the dynamics on 
the energy shell set
by the initial conditions of each trajectory. The TGA, amended by the global 
harmonic approximation for the stability of trajectories, is comparable 
in accuracy to the HK FGA propagator. 

More detailed studies, e.\ g.\ of the convergence properties of the different 
methods, on the choice of an optimal width parameter $\gamma$ for FGA,
and on the applicability of the improved TGA also to unbound problems
are worthwhile topics for future investigations.

\section*{Acknowledgements}
We would like to thank Rick Heller and Eli Pollak for useful discussions.

%
\newpage
\noindent FIG. 1: Comparison of different semiclassical versus quantum 
autocorrelation functions and the semiclassical results for the norm
for the Baranger potential (a) and (c) and the Morse potential
(b) and (d) with the potential parameters given in the text. Full line: 
quantum result, dashed line: HK FGA, dotted line: TGA

\vspace{1cm}
\noindent FIG. 2: The real part of the time dependent width parameter
in the TGA expression: (c) standard $\gamma_1$, 
(b) $\tilde{\gamma}_1=\sqrt{\gamma_1}$, (a)
$\bar{\gamma}_1=\gamma_1^{1/4}$.

\vspace{1cm}
\noindent FIG. 3: The standard and generalized TGA results in the case
of the Baranger potential according to the prefactors in
Fig. \ref{fig:gamma_1}. Full line: quantum result, dotted line: TGA

\vspace{1cm}
\noindent FIG. 4: The global harmonic TGA result compared
to the HK FGA results for the Morse oscillator.
Dashed line: HK FGA, dotted line TGA, and full line: global harmonic TGA

\newpage
\begin{figure}
\begin{center}
\epsfig{file=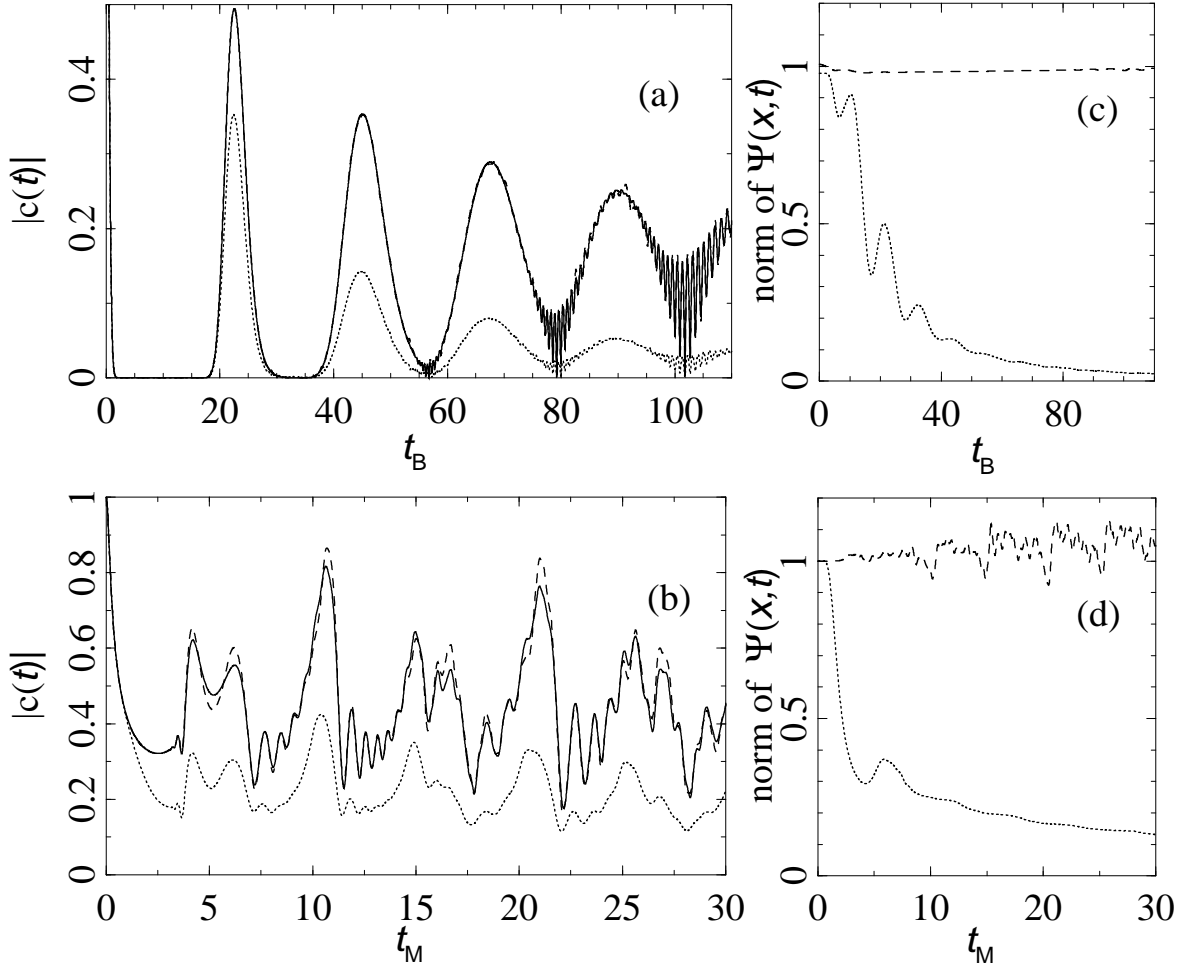,width=13cm,angle=-90}
\caption{\label{fig:ac+norm}
Comparison of different semiclassical versus quantum 
autocorrelation functions and the semiclassical results for the norm
for the Baranger potential (a) and (c) and the Morse potential
(b) and (d) with the potential parameters given in the text. Full line: 
quantum result, dashed line: HK FGA, dotted line: TGA}
\end{center}
\end{figure}

\begin{figure}
\begin{center}
\epsfig{file=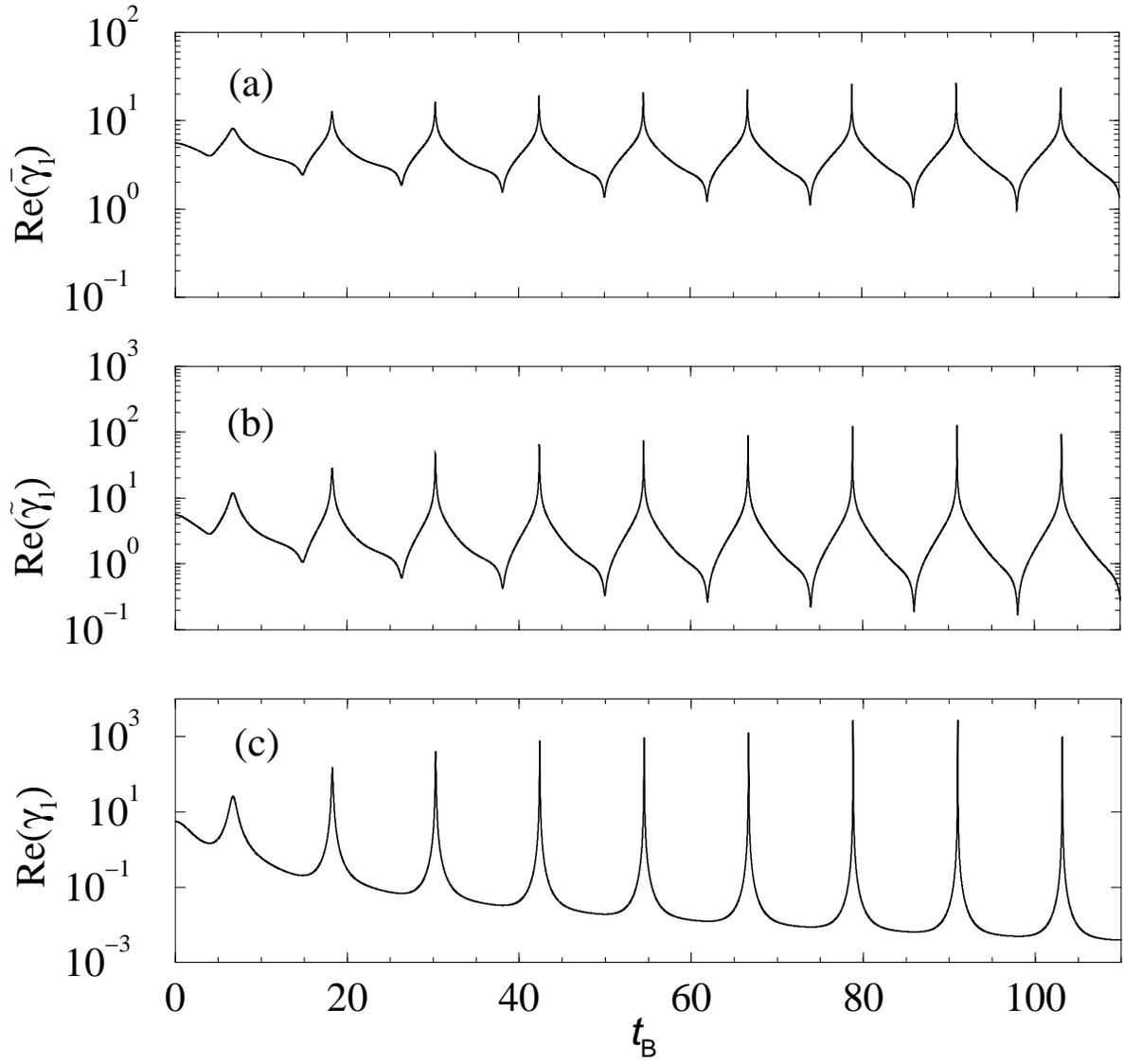,width=15cm,angle=-90}
\caption{\label{fig:gamma_1}
The real part of the time dependent width parameter
in the TGA expression: (c) standard $\gamma_1$, 
(b) $\tilde{\gamma}_1=\sqrt{\gamma_1}$, (a) $\bar{\gamma}_1=\gamma_1^{1/4}$.}
\end{center}
\end{figure}

\begin{figure}
\begin{center}
\epsfig{file=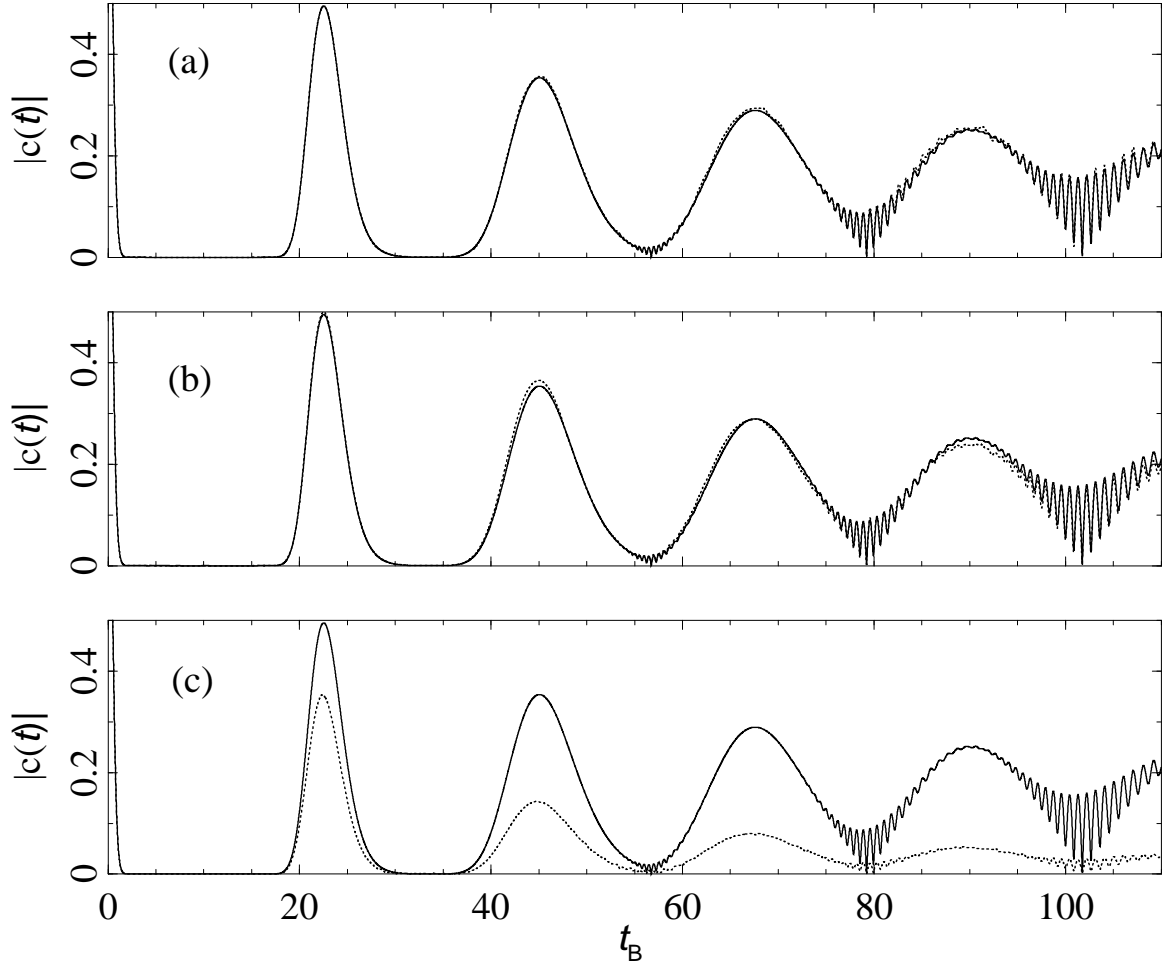,width=13cm,angle=-90}
\caption{\label{fig:bara}
The standard and generalized TGA results in the case
of the Baranger potential according to the prefactors in
Fig. \ref{fig:gamma_1}. Full line: quantum result, dotted line: TGA}
\end{center}
\end{figure}

\begin{figure}
\begin{center}
\epsfig{file=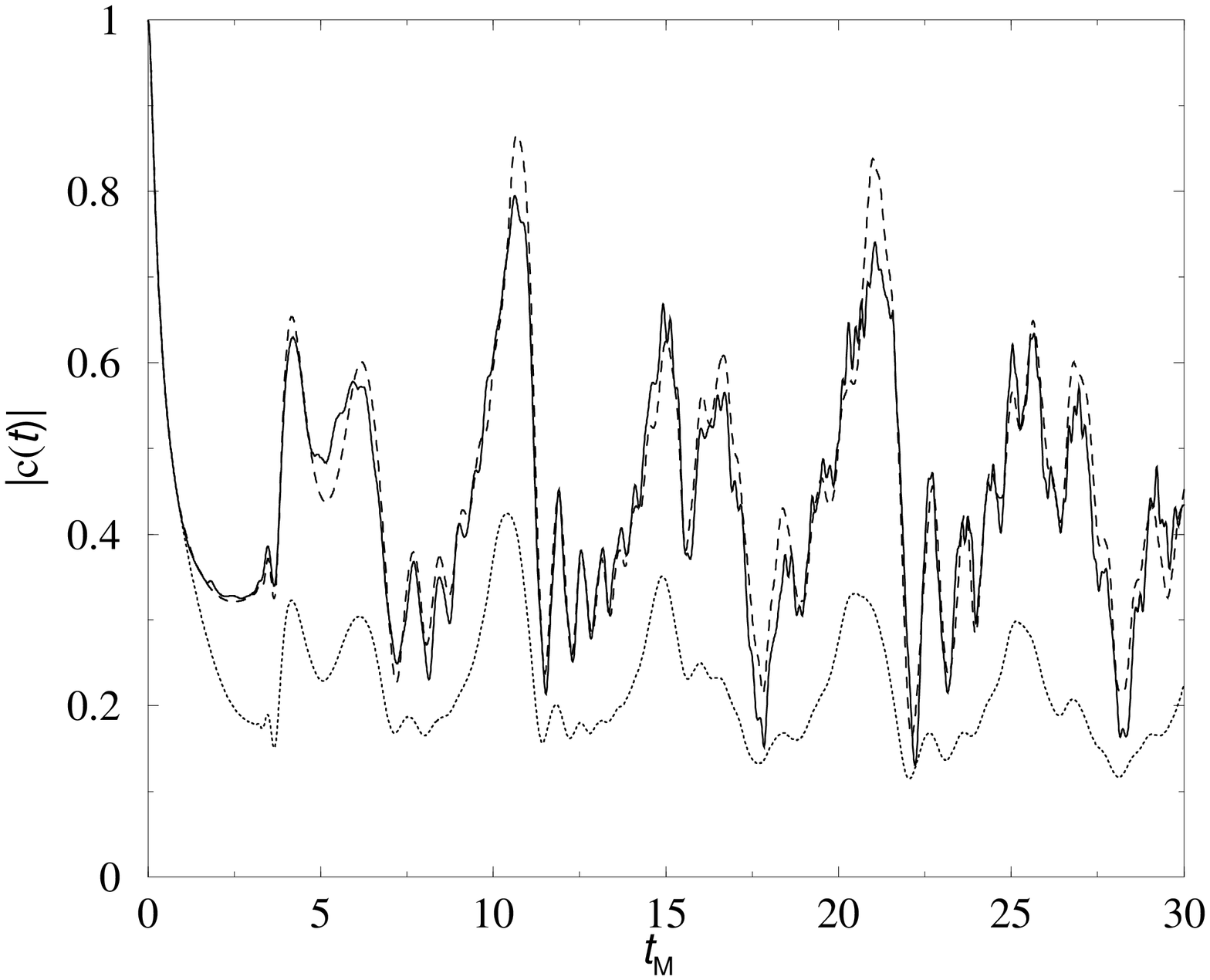,width=13cm,angle=-90}
\caption{\label{fig:morse}
The global harmonic TGA result compared
to the HK FGA results for the Morse oscillator.
Dashed line: HK FGA, dotted line TGA, and full line: global harmonic TGA}
\end{center}
\end{figure}


\begin{thebibliography}{99}
\bibitem{remark2}
$\gamma$ can be also taken complex if the real part is positive;
a slight generalization of the final result is then necessary \cite{Kay1}
\bibitem{remark3}
Especially for TGA this is {\it not} the optimum choice for the 
numerics but there is no
straightforward way to optimze the width parameter and therefore, we stick
to this simple choice.
\end{thebibliography}
\end{document}